\documentclass[journal]{IEEEtran}

\usepackage[T1]{fontenc}
\usepackage{amsmath,amssymb,amsfonts}
\usepackage{algorithmic}
\usepackage{algorithm}
\usepackage{graphicx}
\usepackage{textcomp}
\usepackage{xcolor}
\usepackage{booktabs}
\usepackage{threeparttable}
\usepackage{multirow}
\usepackage{cite}
\usepackage{url}
\usepackage{hyperref}
\usepackage{balance}
\newtheorem{theorem}{Theorem}

\begin{document}

\title{Bernoulli CUSUM and Bayes-Optimal Detection Ceilings for Trust Fraud in Sparse Rating Networks}

\author{Talal~Ashraf~Butt
  \thanks{Manuscript received XXXX; revised XXXX. This work was supported by XXXX.}%
  \thanks{T.~A.~Butt is with Higher Colleges of Technology, Fujairah (e-mail: tbutt@hct.ac.ae).}%
}

\markboth{Preprint}
{Butt: Bernoulli CUSUM and Bayes-Optimal Detection Ceilings}

\maketitle

\begin{abstract}
Sequential trust detection in rating networks relies on continuous observation models that fail on real data. On Bitcoin-OTC, 56\% of ratings take a single value under standard mapping, breaking the distributional assumptions that parametric detectors require. This paper makes three contributions. It derives a Bayes-optimal F1 detection ceiling for per-node sequential detectors using empirically measured observation parameters. At Bitcoin-OTC's median in-degree of 2, this ceiling falls to 0.451 for strategic attacks, explaining why unsupervised methods cluster near $F1 \approx 0.4$. The analysis shows that detector-model matching, not information content, determines performance: binary models retain 86\% of mutual information while enabling exact parametric fit. A dual-regime architecture is presented where Bernoulli CUSUM detects behavioral shifts and triggers asymmetric scoring. Ablation reveals a co-design constraint: the modulation mechanism improves AUC by 0.030 on binary observations but degrades it by 0.094 on continuous observations. The combined system achieves AUC 0.749 on Bitcoin-OTC and 0.796 on Bitcoin-Alpha, beating GaaSTrust on all 8 attacks ($p < 0.003$), with founder-label AUC of 0.999.
\end{abstract}


\section{Introduction}
\label{sec:introduction}

Trust scores in online rating networks serve as the primary defense against adversarial manipulation. When a trader on Bitcoin-OTC receives a sequence of ratings from counterparties, those ratings carry evidence about whether the trader is honest or fraudulent. The natural framework for processing this evidence is sequential change-point detection: monitor the incoming rating stream and flag an alarm when it shifts from the honest baseline. Page's CUSUM~\cite{page1954continuous} is the canonical tool for this task, and Moustakides~\cite{moustakides1986optimal} proved it minimax optimal under exponential-family observations. Two prior efforts have applied CUSUM to trust systems~\cite{liu2010anomaly, li2014quickest}. Both assumed continuous observations drawn from well-behaved distributions.

That assumption has never been tested on real rating data. Bitcoin-OTC is the standard benchmark for trust-based fraud detection, used across graph neural network~\cite{derr2018sgcn, kudo2020gcnext}, iterative scoring~\cite{kumar2018rev2}, and structural feature~\cite{skorupka2024structural} evaluations. The dataset contains 35,592 signed ratings on a scale from $-10$ to $+10$. A standard linear mapping normalizes these to $[0,1]$ for use by continuous detectors. What nobody has reported is what this mapping does to the distribution. Of 35,592 ratings, 56\% equal $+1$, and linear normalization maps these to 0.55 (Figure~\ref{fig:distribution}). A single value carries more than half the probability mass. The Gaussian assumption that continuous CUSUM requires is not approximately wrong. It is categorically wrong.

Continuous CUSUM, calibrated for Gaussian observations, cannot extract usable signal from a distribution that concentrates more than half its mass at a single value. The parametric log-likelihood ratio assumes a smooth density. When the data is effectively discrete, that ratio loses its discriminative power. A simple exponential moving average (GaaSTrust), which makes no distributional assumptions at all, outperforms continuous CUSUM on Bitcoin-OTC across multiple attack types. The problem is not the CUSUM framework. The problem is the observation model fed into it.

This paper makes three contributions.

\textbf{C1. Detection ceiling.} A Bayes-optimal F1 bound is derived for any per-node unsupervised sequential detector operating on Bernoulli observations. On Bitcoin-OTC, where the median in-degree is 2, this ceiling falls to $F1 \leq 0.451$ for strategic attacks. Using empirically measured observation parameters, which differ from attack specifications because honest evaluators independently detect mediocre adversary output, the ceiling explains a decade of stagnation in unsupervised trust detection performance (Section~\ref{sec:ceiling}).

\textbf{C2. Detector-model matching.} This paper documents for the first time that 56\% of Bitcoin-OTC ratings collapse to a single mapped value under standard normalization. Continuous models preserve more mutual information (0.038 bits vs 0.033 bits for strategic attacks), but Gaussian CUSUM cannot extract it from bimodal data. Binary models sacrifice 14\% of MI but match Bernoulli CUSUM's parametric assumptions exactly. Detection performance is governed by this match, not by raw information content (Section~\ref{sec:matching}).

\textbf{C3. Dual-regime Bernoulli CUSUM.} A dual-regime architecture is presented where Bernoulli CUSUM detects behavioral shifts and triggers asymmetric EMA scoring, co-designed with binary observations. Ablation shows binarization contributes +0.015 AUC and CUSUM-triggered modulation contributes +0.030, with a non-decomposable interaction: the same modulation on continuous data degrades performance by 0.094. On Bitcoin-OTC, BernoulliCUSUM reaches AUC 0.749 across 8 attack types, significantly above GaaSTrust's 0.704 ($p < 0.003$). Bitcoin-Alpha yields AUC 0.796, with a founder-label AUC of 0.999 (Section~\ref{sec:experiments}).

Prior work on trust detection and sequential analysis, along with the Bitcoin-OTC benchmark literature, is reviewed in Section~\ref{sec:related}. Section~\ref{sec:ceiling} derives the detection ceiling theorem. The matching diagnosis (Section~\ref{sec:matching}) explains why continuous CUSUM fails on Bitcoin-OTC and motivates the dual-regime Bernoulli design (Section~\ref{sec:method}). Experiments on both Bitcoin datasets validate the approach, and an ablation isolates the contribution of binarization from CUSUM-triggered modulation (Section~\ref{sec:experiments}).

\section{Related Work}
\label{sec:related}

Trust fraud detection draws on three bodies of work: trust-specific scoring methods, sequential change-point detection from statistical process control, and the growing literature that uses Bitcoin-OTC as a benchmark. The common thread across all three is a neglected question, whether the observation model actually matches the data.

\subsection{Trust Detection in Rating Networks}
\label{sec:related:trust}

Bayesian trust modeling began with Bernoulli interactions. In the Beta Reputation System~\cite{josang2002beta}, each interaction feeds a Beta posterior over honest-behavior probability. Network-wide propagation followed with EigenTrust~\cite{kamvar2003eigentrust}, which runs power iteration on the trust matrix. Both methods are static. They compute a single trust score from aggregate evidence and have no mechanism for detecting when a node's behavior changes.

Liu and Sun~\cite{liu2010anomaly} were the first to apply CUSUM to peer-to-peer trust, running a standard Gaussian change-point detector on continuous reputation scores evaluated on synthetic networks with well-separated honest and adversary distributions. Li and Wang~\cite{li2014quickest} extended this to a multi-agent setting, applying a GLR detector under a discrete rating model on synthetic networks with well-separated honest and adversary distributions. Neither paper evaluated whether their assumed rating distribution matches real-world data. On Bitcoin-OTC, the standard continuous mapping produces a distribution that violates the assumptions of every parametric detector tested.

REV2~\cite{kumar2018rev2} jointly estimates user fairness, product goodness, and rating reliability through iterative convergence to a static equilibrium. A strategic adversary who maintains 70\% honest behavior achieves high fairness because the equilibrium reflects average behavior, not temporal shifts. BIRDNEST~\cite{hooi2016birdnest} scores users by the surprise of their rating behavior relative to a population baseline, detecting anomalous raters rather than anomalous targets and relying on batch surprise instead of sequential detection. FRAUDAR~\cite{hooi2016fraudar} identifies fraud through dense subgraph detection, finding groups of users who collectively rate the same products but offering no mechanism for individual behavioral change. Jeong et al.~\cite{jeong2024rater} proposed rater tendency scaling on Bitcoin-OTC, adjusting ratings by per-rater bias. Their work documented rating heterogeneity but did not address the distributional collapse that occurs under linear normalization.

On the GNN side, TrustGNN~\cite{huo2023trustgnn} uses learnable propagation for trust evaluation, and TrustGuard~\cite{wang2024trustguard} targets resilient trust evaluation with dynamicity support. Both are supervised methods that predict trust between node pairs, a fundamentally different task requiring labeled training data that is unavailable in most real-world trust networks.

Every approach listed here either assumes the observation model matches the data or ignores the observation model entirely. None has examined how the structure of the rating distribution affects detector performance.

\subsection{Sequential Detection in Statistical Process Control}
\label{sec:related:spc}

The statistical process control (SPC) community developed Bernoulli CUSUM specifically for monitoring binary outcomes. Optimal threshold selection for detecting shifts in a Bernoulli proportion parameter was derived by Reynolds and Stoumbos~\cite{reynolds1999cusum}. Surgical outcome monitoring provided the first major field application, where Steiner et al.~\cite{steiner2000monitoring} tracked whether a surgeon's mortality rate had shifted from an acceptable baseline. That application shares structural features with trust detection: sparse binary observations per monitored entity, a known in-control rate, and a need for rapid alarm with controlled false positives.

For categorical data, H\"{o}hle~\cite{hohle2010categorical} established that categorical CUSUM outperforms continuous approximations in public health surveillance. Multi-channel settings preserve CUSUM's optimality guarantees under a union rule~\cite{mei2010multichannel}. In the trust-detection setting of this paper, each target node instantiates an independent monitoring channel with categorical observations. Further extensions to non-stationary baselines~\cite{romano2023focus} and high-dimensional streams~\cite{gong2022nncusum} lie beyond the scope of the sparse binary setting considered here.

Despite this mature theoretical foundation, no prior work has applied Bernoulli CUSUM or categorical CUSUM to trust fraud detection. The SPC and trust detection literatures have developed in parallel without cross-pollination.

\subsection{Bitcoin-OTC as Benchmark}
\label{sec:related:bitcoin}

Bitcoin-OTC and Bitcoin-Alpha are the most widely used benchmarks for trust-aware methods on signed networks, introduced by Kumar et al.~\cite{kumar2016bitcoin}. SGCN~\cite{derr2018sgcn} applies balance theory through signed graph convolution to predict edge signs, extended to directed balance theory in GCNEXT~\cite{kudo2020gcnext}. Skorupka et al.~\cite{skorupka2024structural} extracted structural node features and fed them to supervised classifiers, reporting AUC above 0.93 with labeled training data. Islam et al.~\cite{islam2024itrustbd} analyzed trust propagation patterns in Bitcoin networks, and Shadrooh and N{\o}rv{\aa}g~\cite{shadrooh2025datis} proposed GAN-based data augmentation
for node-level trust intensity prediction in incomplete signed networks.

These approaches solve a different problem. Sign prediction assumes stable node identities and asks whether the next edge will be positive or negative. Fraud detection must handle nodes that change behavior over time. The evaluation methodology in this paper reflects this distinction: adversarial agents with known attack strategies are injected onto the real Bitcoin network topology, preserving the original graph structure while controlling the ground truth.

\section{Detection Ceiling}
\label{sec:ceiling}

Any detector's performance is bounded by the information available in the observation stream. A per-node sequential detector observes binary ratings for each target node and must classify that node as honest or adversarial. The central question is how high the F1 score can reach given $n$ observations per node.

Lorden's minimax formulation~\cite{lorden1971procedures} fixes a false-alarm constraint and asks which detection procedure keeps worst-case delay smallest. CUSUM attains that bound whenever the observation distribution sits in the exponential family~\cite{moustakides1986optimal}, the condition this paper establishes for the binary rating model. Multi-stream extensions and finite-sample approximations followed~\cite{tartakovsky2014sequential}. These results characterize detection delay, not classification accuracy. Fraud detection requires a bound on the achievable F1 given a fixed evidence budget.

A Bayes-optimal bound arises under four conditions: (i) observations are i.i.d.\ Bernoulli with parameter $p_0$ under the honest hypothesis and $p_1$ under the adversary hypothesis, (ii) node labels are independent, (iii) the detector uses only per-node evidence (no graph structure), and (iv) the pre-change and post-change parameters are known. Under these conditions, the MAP classifier reaches the F1 ceiling for any per-node detector.

\begin{theorem}[Bayes-Optimal F1 Ceiling]
\label{thm:ceiling}
Given $n$ observations per node, each drawn from $\mathrm{Bernoulli}(p_0)$ under the null hypothesis $H_0$ (honest) or $\mathrm{Bernoulli}(p_1)$ under $H_1$ (adversary), with adversary prior $\pi$, the posterior probability of the adversary hypothesis given total count $S = \sum_{t=1}^{n} X_t$ is
\begin{equation}
P(H_1 \mid S{=}s) = \frac{\pi \cdot p_1^s (1{-}p_1)^{n-s}}%
  {\pi \cdot p_1^s (1{-}p_1)^{n-s} + (1{-}\pi) \cdot p_0^s (1{-}p_0)^{n-s}}.
\label{eq:posterior}
\end{equation}
The MAP classifier declares adversary when $P(H_1 \mid S) > 0.5$, and the resulting F1 score is the ceiling for any per-node detector whose decision function depends on the observations only through the count sufficient statistic $S = \sum_{i} X_i$.
\end{theorem}

The ceiling depends on empirical observation parameters, not the attack design specification. Table~\ref{tab:attacks} distinguishes between two quantities: the specification $p_1$ (what the adversary produces) and the empirical $p_1$ (what the adversary receives, as measured after injection). Honest evaluators independently react to adversary behavior. The divergence between specification and measurement can be large. For bad\_mouth (specification $p_1 = 0.85$), evaluators detect mediocre output and downrate adversaries. The empirical received positive rate drops to $p_1 = 0.503$. This community amplification increases the KL divergence from 0.003 nats (specification parameters) to 0.111 nats (empirical), a 37-fold increase in per-observation discriminability. All ceiling computations in this paper use empirical parameters.

\begin{figure}[!t]
\centering
\includegraphics[width=\columnwidth]{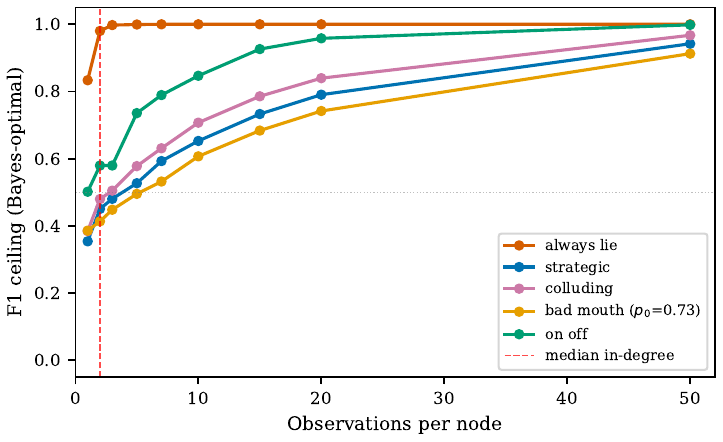}
\caption{Bayes-optimal F1 ceiling computed with empirically measured observation parameters (Table~\ref{tab:attacks}). Bad\_mouth uses $p_0 = 0.731$ (honest positive rate reduced by adversary downrating). The dashed vertical line marks $n = 2$ (median in-degree of Bitcoin-OTC).}
\label{fig:ceiling}
\end{figure}

Figure~\ref{fig:ceiling} plots this ceiling as a function of $n$ for five attack types using empirical observation parameters (Table~\ref{tab:attacks}). The dashed vertical line marks $n = 2$, the median in-degree of Bitcoin-OTC. At this observation level, the ceiling sits at 0.451 for strategic attacks. For bad\_mouth, the empirical parameters ($p_0 = 0.731$, $p_1 = 0.503$) produce a ceiling of 0.414, substantially higher than the specification-based ceiling of 0.333. Community amplification makes bad\_mouth more detectable than the attack design suggests.

Unsupervised methods on Bitcoin-OTC have clustered at $F1 \approx 0.3$--$0.4$ for years. The constraint is not algorithmic. At median in-degree 2, the per-node Bayesian error rate dominates detection performance regardless of which method is applied. The situation parallels the sparse mixture detection problem studied by Donoho and Jin~\cite{donoho2004higher}, where sparse signal and low per-observation discriminability impose hard information-theoretic limits even on optimal detectors. Structured observations can improve detection beyond naive per-stream bounds~\cite{xie2021sequential}.

Table~\ref{tab:stratified} stratifies detection efficiency by in-degree. BernoulliCUSUM achieves 95.1\% of the Bayes-optimal ceiling for nodes with median in-degree 2. Efficiency decreases for well-observed nodes (76.2\% at median in-degree 19), where the ceiling rises faster than detection improves.

\section{Detector-Model Matching}
\label{sec:matching}

The ceiling theorem establishes what is achievable. Existing continuous detectors fall short of that bound because of the observation model, not the detection algorithm.

\begin{figure}[!t]
\centering
\includegraphics[width=\columnwidth]{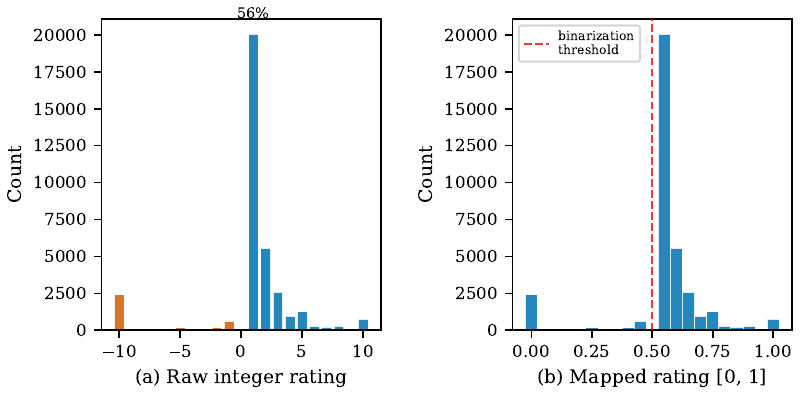}
\caption{Bitcoin-OTC rating distribution. (a)~Raw integer histogram showing the dominant spike at $+1$ (56\% of all ratings). (b)~Standard linear mapping to $[0,1]$ produces a point mass at 0.55, violating the continuous distributional assumptions of Gaussian CUSUM.}
\label{fig:distribution}
\end{figure}

Figure~\ref{fig:distribution}(a) shows the raw rating distribution of Bitcoin-OTC. Of the 35,592 ratings, 56\% take the value $+1$. The remaining mass spreads thinly across the $[-10, +10]$ integer range, with a secondary concentration at $-10$. Under the standard linear mapping $r \mapsto (r + 10)/20$, the $+1$ spike lands at 0.55 (Figure~\ref{fig:distribution}(b)). One value holds more than half the probability mass.

A natural objection is that binarization discards information. Mutual information~\eqref{eq:mi} and KL divergence~\eqref{eq:kl} between honest and adversary observation distributions test this directly.

The mutual information between the observation and the hypothesis is
\begin{equation}
I(X; H) = H(X) - H(X \mid H),
\label{eq:mi}
\end{equation}
where $H(X \mid H) = (1-\pi) \cdot H(\mathrm{Ber}(p_0)) + \pi \cdot H(\mathrm{Ber}(p_1))$.

The KL divergence between adversary and honest observation distributions is
\begin{equation}
D_{\mathrm{KL}}(\mathrm{Ber}(p_1) \| \mathrm{Ber}(p_0)) = p_1 \ln\frac{p_1}{p_0} + (1 - p_1) \ln\frac{1-p_1}{1-p_0}.
\label{eq:kl}
\end{equation}

For strategic attacks, the continuous model retains 0.038 bits of mutual information. The binary model retains 0.033 bits, sacrificing 14\% of MI. The continuous model carries more raw signal. It should win.

It does not. Gaussian CUSUM's log-likelihood ratio assumes a unimodal density with known mean and variance. When 56\% of the data mass concentrates at a single value, that assumption collapses. Because the Gaussian density assigns vanishing probability to an observed point mass, the LLR cannot accumulate in any consistent direction. Detection delay grows sharply as the assumed distribution diverges from the true data-generating process~\cite{gong2022nncusum}. On discrete or heavily quantised data, the degradation is more severe. A GLR-based binning statistic that partitions the sample space consistently outperforms Gaussian CUSUM when the post-change distribution is not fully specified~\cite{lau2019binning}.

Bernoulli CUSUM faces no such problem. Its LLR takes only two values, one for each outcome $X_t \in \{0,1\}$. The Bernoulli family lies in the exponential family, and Moustakides' optimality guarantee~\cite{moustakides1986optimal} activates in full.

The performance gap between continuous and Bernoulli CUSUM on bimodal data is driven by extraction efficiency, not information loss. The binary model retains 86\% of mutual information. Bernoulli CUSUM's parametric assumptions are exactly satisfied. Gaussian CUSUM's are not even approximately satisfied. Detection performance is governed by the match between the observation model and the detector's distributional assumptions~\cite{cover2006elements}, not by the raw information content of the observations.

\begin{figure}[!t]
\centering
\includegraphics[width=\columnwidth]{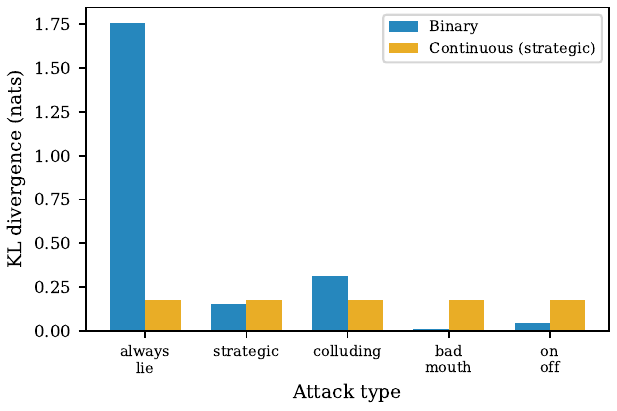}
\caption{KL divergence between adversary and honest observation distributions under binary and continuous models, across five attack types. Binary KL dominates when $p_1$ is far from $p_0$ (always\_lie, colluding). Continuous KL dominates when $p_1 \approx p_0$ (bad\_mouth, on\_off).}
\label{fig:kl}
\end{figure}

Figure~\ref{fig:kl} extends this analysis across five attack types. For always\_lie ($p_1 = 0.10$), binary KL divergence reaches 1.75 nats, nearly $10\times$ the continuous value of 0.18 nats. Binarization concentrates the full separation between $p_0 = 0.90$ and $p_1 = 0.10$ into a single binary contrast that Bernoulli CUSUM extracts completely. For colluding ($p_1 = 0.60$), the binary advantage is smaller at 0.31 vs 0.18 nats. A crossover occurs at bad\_mouth and on\_off, where $p_1$ approaches $p_0$ and the continuous model retains more discriminability. These are exactly the attacks where BernoulliCUSUM's per-attack advantage shrinks (Table~\ref{tab:full_auc}). The KL structure predicts the empirical performance ordering.

\section{Method: Dual-Regime Bernoulli CUSUM}
\label{sec:method}

\begin{figure*}[!t]
\centering
\includegraphics[width=\textwidth]{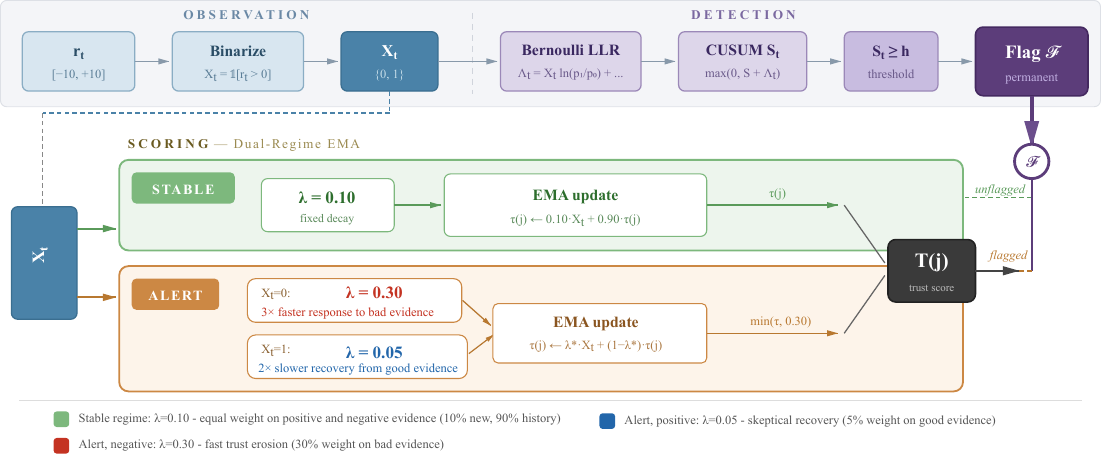}
\caption{Dual-regime BernoulliCUSUM pipeline. Raw ratings are binarised into $\{0,1\}$ by the Observation layer. The Detection layer runs Bernoulli CUSUM and permanently flags suspect nodes when $S_t \geq h$. Unflagged nodes receive a stable EMA update at $\lambda = 0.10$. Flagged nodes switch to asymmetric decay ($\lambda_{\text{neg}} = 0.30$, $\lambda_{\text{pos}} = 0.05$) with output capped at 0.30.}
\label{fig:pipeline}
\end{figure*}

Each rating is binarised, an honest baseline is calibrated on an initial window, Bernoulli CUSUM runs on the resulting binary stream, and the alarm feeds into a dual-regime EMA trust score. Algorithm~\ref{alg:bcusum} and Figure~\ref{fig:pipeline} present the procedure. CUSUM flagging drives the regime switch that reweights the EMA update.

\paragraph{Binarization.}
Each incoming rating $r_t$ for target node $j_t$ collapses to the sign indicator $X_t = \mathbf{1}[r_t > 0]$. A positive rating encodes as 1, and any rating at or below zero encodes as 0. This mapping discards magnitude information but matches the distributional structure of Bitcoin-OTC, where the distinction between positive and negative carries more signal than the distinction between $+1$ and $+5$ (Section~\ref{sec:matching}).

\paragraph{Phase I calibration.}
The in-control positive rate $p_0$ is estimated from an initial window of ratings under the assumption that the system starts in a predominantly honest state. On Bitcoin-OTC, the empirical positive rate is $p_0 = 0.90$. The post-change parameter is set to $p_1 = p_0 - 0.20 = 0.70$, calibrated against the strategic attack.

\paragraph{Bernoulli CUSUM monitoring.}
For each target node $j$, the detector maintains a CUSUM statistic $S_j$. Under Bernoulli observations, each arrival contributes an LLR increment
\begin{equation}
\Lambda_t = X_t \cdot \ln\frac{p_1}{p_0} + (1 - X_t) \cdot \ln\frac{1-p_1}{1-p_0}.
\label{eq:llr}
\end{equation}

The increments accumulate through Page's non-negative recursion,
\begin{equation}
S_t = \max(0,\; S_{t-1} + \Lambda_t), \quad S_0 = 0.
\label{eq:cusum}
\end{equation}

An alarm is raised the first time $S_t$ crosses the threshold $h = 3.0$, and the resulting flag never clears. For exponential-family observations, this recursion minimises the essential supremum of detection delay subject to a false-alarm constraint on mean time~\cite{moustakides1986optimal}. The Bernoulli family satisfies the exponential-family condition exactly, so the minimax optimality guarantee holds without approximation.

\paragraph{Dual-regime EMA scoring.}
The CUSUM flag triggers a regime switch in the EMA trust score. In the stable regime (unflagged), the EMA uses a fixed decay $\lambda = 0.10$, weighting each new observation at 10\% and retaining 90\% of history. Once CUSUM detects a regime change ($S \geq h$), the EMA switches to asymmetric parameters: negative evidence is weighted $3\times$ more heavily ($\lambda_{\text{neg}} = 0.30$) while positive evidence recovery is halved ($\lambda_{\text{pos}} = 0.05$). A detected adversary should not recover trust quickly from occasional positive interactions. At output, flagged nodes receive a score cap of 0.30.

\paragraph{Co-design constraint.}
The dual-regime mechanism requires binary observations. Applied to continuous $[0,1]$ observations, the asymmetric lambdas amplify point-mass noise at 0.55 instead of genuine positive/negative signal, degrading AUC from 0.704 to 0.610 (Section~\ref{sec:exp:ablation}, Table~\ref{tab:ablation}). Binary observations create clean $\{0, 1\}$ separation that the dual-regime modulation amplifies. The observation model and detection architecture are not independently substitutable.

\paragraph{Parameter summary.}
Throughout, the parameters are fixed at $p_0 = 0.90$, $p_1 = 0.70$, $h = 3.0$, $\lambda = 0.10$, $\lambda_{\text{neg}} = 0.30$, $\lambda_{\text{pos}} = 0.05$, and a score cap of 0.30, without any per-attack adjustment.

\paragraph{Complexity.}
A single temporal pass over the edge list suffices, with every edge triggering a constant-time update. Time scales as $O(|E|)$ in the number of ratings, and memory is $O(|V|)$ per target node. On Bitcoin-OTC (35,592 edges, 5,881 nodes), a single CPU core finishes in under a second.

\paragraph{Regime selection.}
BernoulliCUSUM is not universally superior. On data with balanced, high-variance, or uniform rating distributions, continuous CUSUM outperforms it (Section~\ref{sec:exp:regime}). If more than 50\% of observations map to a single value under the standard normalization, deploy Bernoulli CUSUM. Otherwise, deploy continuous CUSUM.

\begin{algorithm}[!t]
\caption{Dual-Regime Bernoulli CUSUM Trust Detection}
\label{alg:bcusum}
\begin{algorithmic}[1]
\REQUIRE Rating stream $\{(i_t, j_t, r_t)\}_{t=1}^{T}$; parameters $p_0$, $p_1$, $h$, $\lambda$, $\lambda_{\text{neg}}$, $\lambda_{\text{pos}}$
\ENSURE Trust scores $\tau(j)$ for each target node $j$
\STATE \textbf{Phase I:} Estimate $\hat{p}_0$ from initial ratings
\FOR{each interaction $(i_t, j_t, r_t)$ at time $t$}
  \STATE Binarize: $X_t \gets \mathbf{1}[r_t > 0]$
  \STATE Compute LLR: $\Lambda_t \gets X_t \ln\frac{p_1}{p_0} + (1{-}X_t) \ln\frac{1{-}p_1}{1{-}p_0}$
  \STATE Update CUSUM: $S_{j_t} \gets \max(0,\; S_{j_t} + \Lambda_t)$
  \IF{$S_{j_t} \geq h$}
    \STATE Flag $j_t$ (permanent)
  \ENDIF
  \IF{$j_t$ is flagged}
    \STATE $\lambda^* \gets \lambda_{\text{neg}}$ if $X_t = 0$, else $\lambda_{\text{pos}}$ \COMMENT{Asymmetric}
  \ELSE
    \STATE $\lambda^* \gets \lambda$ \COMMENT{Stable regime}
  \ENDIF
  \STATE $\tau(j_t) \gets \lambda^* \cdot X_t + (1 - \lambda^*) \cdot \tau(j_t)$ \COMMENT{EMA update}
\ENDFOR
\STATE \textbf{Output:} $\tau(j) \gets \min(\tau(j),\, 0.30)$ if $j$ flagged, else $\tau(j)$
\end{algorithmic}
\end{algorithm}

\section{Experiments}
\label{sec:experiments}

BernoulliCUSUM is evaluated on two real-world Bitcoin trust networks against four baselines under eight attack types. The analysis covers distributional regime crossover and decomposes the contribution of each architectural component through ablation.

\subsection{Experimental Setup}
\label{sec:exp:setup}

\paragraph{Datasets.}
Bitcoin-OTC contains 5,881 nodes and 35,592 directed, signed, weighted edges representing trader-to-trader trust ratings on the $[-10, +10]$ integer scale. The median in-degree is 2 and 90.0\% of ratings are positive. Bitcoin-Alpha contains 3,783 nodes and 24,186 edges with 93.6\% positive ratings and median in-degree 2.

\paragraph{Attack injection.}
Adversarial agents are injected onto the real network topology, replacing 20\% of nodes with adversaries following one of eight attack strategies.\footnote{Under static attack injection, sybil and whitewash produce equivalent behavior. This paper reports sybil. Whitewash requires dynamic re-entry simulation beyond this evaluation's scope.} Table~\ref{tab:attacks} describes each strategy with both specification and empirical observation parameters. Each configuration is run with 10 random seeds to control adversary placement.

\begin{table}[!t]
\caption{Attack strategies with specification and empirical observation
parameters. Ceiling computed using empirical values.}
\label{tab:attacks}
\centering
\small
\resizebox{\columnwidth}{!}{%
\begin{threeparttable}
\begin{tabular}{lcccc}
\toprule
Attack & Behavior & Spec $p_1$ & Emp.\ $p_1$ & Emp.\ $p_0$ \\
\midrule
always\_lie   & All adversarial      & 0.10 & 0.000 & 0.900 \\
strategic     & 70\% honest          & 0.70 & 0.699 & 0.900 \\
colluding     & 60\% adversarial     & 0.60 & 0.667 & 0.900 \\
bad\_mouth    & Mediocre; downrates  & 0.85 & 0.503 & 0.731$^\dagger$ \\
ballot\_stuff & Mediocre; inflates   & 0.85 & 0.585 & 0.900 \\
on\_off       & Phase alternation    & 0.80 & 0.532 & 0.900 \\
oscillation   & Sinusoidal quality   & 0.75 & 0.498 & 0.900 \\
sybil         & Mediocre; colluding  & 0.85 & 0.597 & 0.900 \\
\bottomrule
\end{tabular}
\begin{tablenotes}[flushleft]
\footnotesize
\item Spec $p_1$: attack design parameter (output).
\item Emp.\ $p_1$: measured positive rate received by adversary nodes.
\item[$\dagger$] Bad\_mouth adversaries reduce honest $p_0$ from 0.900 to 0.731.
\end{tablenotes}
\end{threeparttable}
}
\end{table}

\paragraph{Baselines.}
Five methods are compared. BernoulliCUSUM is the proposed method. GaaSTrust is an exponential moving average baseline with no distributional assumptions. BTGAggDA-Cal is a continuous CUSUM variant calibrated on the same data. SimpleMean computes the arithmetic mean of raw ratings per node. REV2~\cite{kumar2018rev2} is the iterative fairness-goodness-reliability algorithm.

\paragraph{Metrics.}
AUC-ROC is the primary metric. It measures ranking quality across all thresholds and is independent of score scale. F1 at the default threshold is reported alongside the detection ceiling from Section~\ref{sec:ceiling} to contextualize absolute F1 values. Statistical significance is assessed via bootstrap confidence intervals (10,000 resamples) with Holm--Bonferroni correction. Effect sizes are reported as Cliff's $\delta$.

\subsection{Bitcoin-OTC Results}
\label{sec:exp:otc}

Table~\ref{tab:main} reports mean AUC across 8 attacks and 10 seeds. BernoulliCUSUM achieves 0.749, followed by GaaSTrust at 0.704, BTGAggDA-Cal at 0.562, SimpleMean at 0.558, and REV2 at 0.554.

\begin{figure*}[!t]
\centering
\includegraphics[width=\textwidth]{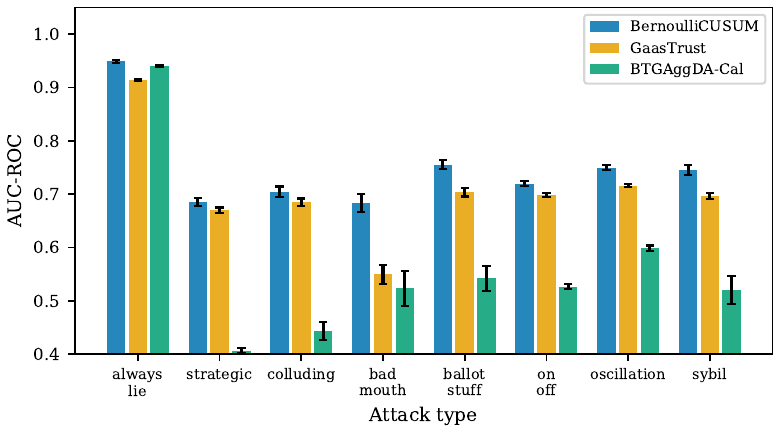}
\caption{Per-attack AUC on Bitcoin-OTC for BernoulliCUSUM, GaaSTrust, and BTGAggDA-Cal. Error bars show 95\% bootstrap confidence intervals over 10 seeds. BernoulliCUSUM achieves the highest AUC on all 8 attacks ($p < 0.003$, Holm--Bonferroni corrected).}
\label{fig:auc}
\end{figure*}

Figure~\ref{fig:auc} breaks this down by attack type. BernoulliCUSUM achieves the highest AUC on all 8 attacks (Holm--Bonferroni corrected; seven at $p < 0.0001$, strategic at $p = 0.003$). Against always\_lie, the gap is +0.034 over GaaSTrust (0.949 vs 0.915, Cliff's $\delta = 1.000$). Against strategic, the gap narrows to +0.016 (0.685 vs 0.670, Cliff's $\delta = 0.700$). Bad\_mouth produces the largest single-attack margin at +0.134 (0.683 vs 0.549), where the adversary's negative ratings toward honest nodes create a clear binary signature.

Under attack injection, REV2's iterative convergence assigns high fairness to adversaries who maintain a majority of honest interactions, yielding only AUC 0.554. BTGAggDA-Cal also underperforms at 0.562, confirming that the distributional mismatch diagnosed in Section~\ref{sec:matching} translates directly to detection failure.

Table~\ref{tab:full_auc} provides the full per-attack comparison between BernoulliCUSUM and GaaSTrust. All eight attacks show statistically significant improvements with large effect sizes (Cliff's $\delta \geq 0.620$). The tightest margin is strategic (+0.016), where the adversary's 70\% honest behavior makes the binary signal weakest. The widest margin is bad\_mouth (+0.134), where adverse ratings toward honest nodes produce a measurable drop in the positive rate.

\begin{table}[!t]
\caption{Detection performance on Bitcoin trust networks (mean AUC-ROC, 10 seeds).}
\label{tab:main}
\centering
\begin{tabular}{lccc}
\toprule
 & Bitcoin-OTC & Bitcoin-Alpha & Founder AUC \\
\midrule
BernoulliCUSUM & \textbf{0.749} & \textbf{0.796} & 0.999 \\
GaaSTrust & 0.704 & 0.748 & 0.998 \\
BTGAggDA-Cal & 0.562 & 0.612 & \textbf{1.000} \\
SimpleMean & 0.558 & --- & 1.000 \\
REV2 & 0.554 & --- & 0.633 \\
\bottomrule
\end{tabular}
\end{table}

\begin{table*}[!t]
\caption{Per-attack AUC: BernoulliCUSUM vs GaaSTrust (Bitcoin-OTC, 10 seeds). All comparisons significant after Holm--Bonferroni correction; seven at $p < 0.0001$, strategic at $p = 0.003$.}
\label{tab:full_auc}
\centering
\begin{tabular}{lcccccc}
\toprule
Attack & BernoulliCUSUM & GaaSTrust & $\Delta$AUC & 95\% CI & $p$-value & Cliff's $\delta$ \\
\midrule
always\_lie & 0.949 & 0.915 & +0.034 & [+0.030, +0.038] & $<$0.0001 & +1.000 \\
strategic & 0.685 & 0.670 & +0.016 & [+0.005, +0.026] & 0.003 & +0.700 \\
colluding & 0.704 & 0.685 & +0.019 & [+0.008, +0.032] & $<$0.0001 & +0.620 \\
bad\_mouth & 0.683 & 0.549 & +0.134 & [+0.125, +0.143] & $<$0.0001 & +1.000 \\
ballot\_stuff & 0.755 & 0.704 & +0.052 & [+0.044, +0.060] & $<$0.0001 & +1.000 \\
on\_off & 0.720 & 0.698 & +0.021 & [+0.019, +0.026] & $<$0.0001 & +0.980 \\
oscillation & 0.750 & 0.716 & +0.034 & [+0.029, +0.038] & $<$0.0001 & +1.000 \\
sybil & 0.745 & 0.696 & +0.049 & [+0.038, +0.059] & $<$0.0001 & +1.000 \\
\bottomrule
\end{tabular}
\end{table*}

On the founder-label evaluation (organic labels from the Bitcoin-OTC dataset, without synthetic attack injection), BernoulliCUSUM achieves AUC 0.999. These labels reflect community consensus about early platform participants. The near-perfect score indicates that the EMA scoring recovers the organic reputation structure of the network.

\begin{table}[!t]
\caption{Detection efficiency by in-degree bin (mean across 8 attacks,
empirical observation parameters, bin-specific thresholds).}
\label{tab:stratified}
\centering
\begin{tabular}{lcccc}
\toprule
In-degree bin & Med.\ $n$ & Ceiling & Achieved F1 & Efficiency \\
\midrule
$n{=}1$ & 1 & 0.487 & 0.494 & 101.6\% \\
$n{=}2$ & 2 & 0.573 & 0.545 & 95.1\% \\
$n{=}3$--$5$ & 4 & 0.640 & 0.576 & 90.0\% \\
$n{=}6$--$10$ & 7 & 0.723 & 0.589 & 81.4\% \\
$n{>}10$ & 19 & 0.883 & 0.672 & 76.2\% \\
\bottomrule
\end{tabular}
\end{table}

\subsection{Bitcoin-Alpha Results}
\label{sec:exp:alpha}

Bitcoin-Alpha confirms generalization. BernoulliCUSUM achieves AUC 0.796 versus GaaSTrust at 0.748 (Table~\ref{tab:main}), with consistent improvement across all 8 attack types. Bad\_mouth shows the largest margin and always\_lie the smallest, mirroring the Bitcoin-OTC ordering. Strategic and on\_off yield the tightest margins, consistent with the pattern on OTC where these attacks produce the weakest binary signal. Bitcoin-Alpha's higher positive rate (93.6\% vs 90.0\%) concentrates the honest baseline further, which amplifies the binary contrast when an adversary's positive rate drops.

\subsection{Regime Crossover Analysis}
\label{sec:exp:regime}

\begin{figure*}[!t]
\centering
\includegraphics[width=\textwidth]{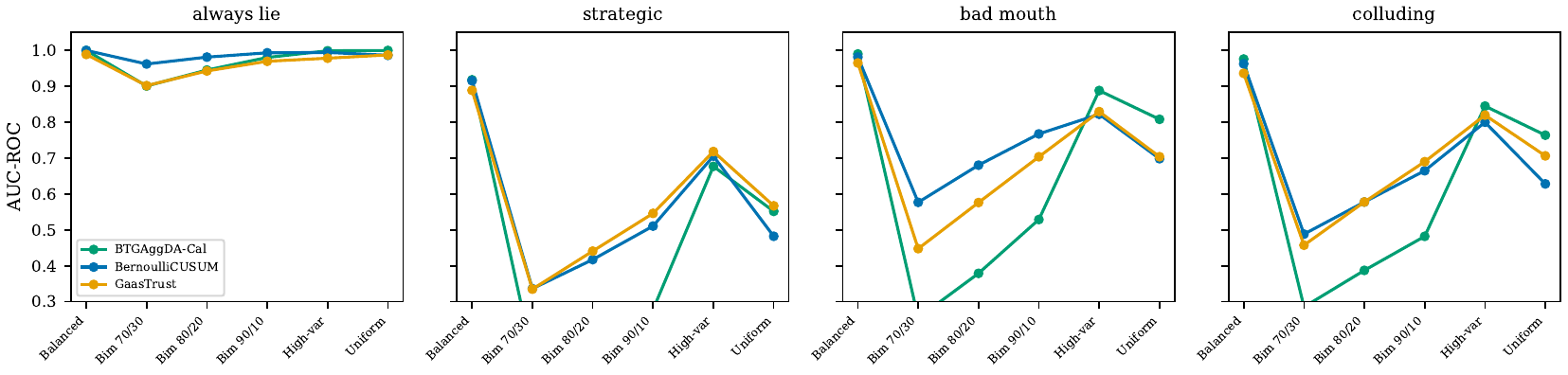}
\caption{Regime crossover across six distributional profiles and four attack types. BernoulliCUSUM dominates on bimodal data with $\geq$80\% concentration at a single value. Continuous CUSUM (BTGAggDA-Cal) outperforms on balanced and uniform profiles. GaaSTrust remains competitive on intermediate regimes.}
\label{fig:regime}
\end{figure*}

BernoulliCUSUM is not the right tool for every distribution. Figure~\ref{fig:regime} tests all three detectors across six distributional profiles: Balanced, Bimodal 70/30, Bimodal 80/20, Bimodal 90/10, High-variance, and Uniform. Each profile is evaluated on four attack types.

On Balanced, High-variance, and Uniform profiles, BTGAggDA-Cal (continuous CUSUM) dominates. These distributions have sufficient spread for Gaussian assumptions to hold. On Bimodal 80/20, BernoulliCUSUM wins across all four attacks, matching the regime of Bitcoin-OTC. Intermediate regimes are messier: on Bimodal 70/30, GaaSTrust is competitive with both CUSUM variants.

\subsection{Ablation and Co-Design Analysis}
\label{sec:exp:ablation}

Table~\ref{tab:ablation} decomposes the AUC improvement over GaaSTrust. Binarization alone (binary EMA, no CUSUM) provides +0.015 AUC. Dual-regime modulation on binary data adds +0.030 over the binary EMA baseline, for a combined +0.045. The raw CUSUM statistic as a stand-alone score achieves 0.720, comparable to binary EMA (0.719), confirming that both mechanisms extract similar information from the binary stream.

\begin{figure}[!t]
\centering
\includegraphics[width=\columnwidth]{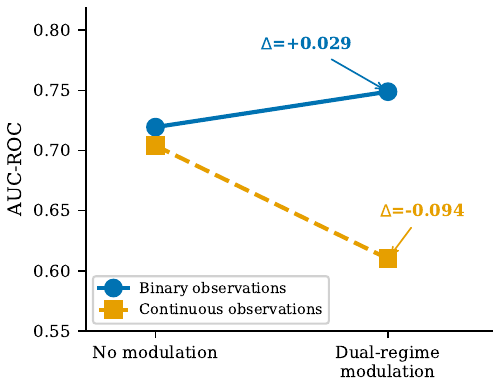}
\caption{Co-design interaction between observation model and dual-regime modulation. Binary observations benefit from modulation ($\Delta = +0.030$). Continuous observations are degraded ($\Delta = -0.094$). The crossing pattern indicates a non-decomposable interaction.}
\label{fig:codesign}
\end{figure}

The critical finding is the co-design constraint (Figure~\ref{fig:codesign}). The same lambda modulation applied to continuous $[0,1]$ observations degrades AUC by 0.094, from 0.704 (GaaSTrust baseline) to 0.610. The interaction effect (+0.124 AUC swing between the continuous and binary modulation deltas in Table~\ref{tab:ablation}) exceeds either component's individual contribution. The observation model and detection architecture are not independently substitutable.

Beyond ranking quality, CUSUM provides something the EMA cannot: a formal detection delay guarantee. Among adversary nodes that accumulate enough observations for the statistic to reach threshold, CUSUM flags them in a median of 3 observations (on\_off) to 18 observations (bad\_mouth). On\_off adversaries trigger fastest because their adversarial phases produce concentrated bursts of negative ratings. At Bitcoin-OTC's sparse in-degree distribution, 19--59\% of adversary nodes receive enough ratings for CUSUM to fire, depending on attack type. For the remaining sparse nodes, the EMA's base scoring provides ranking quality without CUSUM intervention. The dual-regime architecture thus splits the detection task: CUSUM handles well-observed nodes with formal optimality, and the EMA handles sparse nodes with best-effort ranking.

\begin{table}[!t]
\caption{Ablation decomposition: contribution of binarization and
dual-regime modulation to AUC improvement (Bitcoin-OTC, mean across
8 attacks, 10 seeds).}
\label{tab:ablation}
\centering
\begin{tabular}{lcc}
\toprule
Variant & AUC & $\Delta$ vs GaaSTrust \\
\midrule
GaaSTrust (continuous EMA) & 0.704 & --- \\
Binary EMA only & 0.719 & $+0.015$ \\
CUSUM score only & 0.720 & $+0.016$ \\
Continuous EMA + modulation & 0.610 & $-0.094$ \\
\textbf{BernoulliCUSUM (combined)} & \textbf{0.749} & $+$\textbf{0.045} \\
\bottomrule
\end{tabular}
\end{table}

\subsection{Rater-Relative Deviation}
\label{sec:exp:deviation}

An alternative to global binarization is rater-relative deviation: compute each rater's personal mean, then binarize based on whether a given rating deviates from that personal baseline. The intuition is that a consistently positive rater who gives a negative rating carries more signal than one whose ratings already vary.

On Bitcoin-OTC, this approach fails. Of 4,814 unique raters, only 45.7\% have three or more ratings. Per-rater baselines are unreliable at this sparsity. The standard deviation of rater means is 2.140 on the raw $[-10, +10]$ scale (0.117 on the mapped $[0,1]$ scale). Raters are homogeneous. Most rate positively and most rate similarly. Deviation-based binarization adds estimation noise without adding discriminative power, and the resulting detector achieves AUC 0.596, well below BernoulliCUSUM's 0.749.

A separate per-evaluator monitoring stream (Stream~B) that runs Bernoulli CUSUM on each rater's outgoing ratings was also tested. Stream~B improves bad\_mouth detection marginally but reduces founder-label AUC from 0.999 to 0.971. The trade-off between evaluator filtering and organic label recovery favors the simpler single-stream design.

The critical design choice is the distributional form of the observation model (continuous vs binary), not the reference point used for binarization (global threshold vs personal baseline). On data where raters are homogeneous, personalization adds noise without adding signal.

\section{Discussion}
\label{sec:discussion}

Three questions emerge from the experimental results: why AUC is the primary metric instead of F1, where BernoulliCUSUM fails, and what directions follow.

\subsection{Why AUC, Not F1}
\label{sec:disc:auc}

SimpleMean achieves F1 = 0.400 at the default threshold on Bitcoin-OTC, higher than BernoulliCUSUM's F1 = 0.309. At its own optimal threshold, GaaSTrust reaches 0.524, above BernoulliCUSUM's 0.496. Both comparisons mislead for the same structural reason.

SimpleMean operates on raw integer ratings. Its trust scores spread across a wide range. BernoulliCUSUM's EMA scoring compresses scores into a narrow band near the positive rate. The wider spread gives SimpleMean a favorable threshold-dependent F1 despite far worse ranking quality: its AUC is 0.558, nearly 0.2 below BernoulliCUSUM's 0.749. The F1 advantage reflects score spread, not detection quality.

A trust system in deployment ranks nodes and presents the ranking to a human operator or downstream process. It does not make binary accept/reject decisions at a fixed threshold. AUC measures exactly this ranking quality, independent of score scale and threshold choice. The detection ceiling from Section~\ref{sec:ceiling} further contextualizes F1: at median in-degree 2, the Bayes-optimal ceiling stays near or below 0.45 for most attacks. Comparing methods on F1 without ceiling context attributes informational limits to algorithmic failure.

\subsection{Limitations}
\label{sec:disc:limitations}

On balanced continuous distributions where no single value dominates, BernoulliCUSUM underperforms continuous CUSUM by a wide margin. The regime crossover analysis (Figure~\ref{fig:regime}) shows this directly: on the Balanced profile, BernoulliCUSUM drops to near-chance on strategic attacks while continuous CUSUM achieves AUC above 0.90. Binarization discards the distributional structure that continuous detectors exploit.

The co-design constraint means that deploying the dual-regime architecture on continuous data without binarization degrades performance below the non-modulated baseline. The evaluation uses static attack injection, where every injected agent follows a fixed strategy from start to end. An adaptive adversary that monitors its own trust score and throttles behavior to stay below detection thresholds would pose a harder problem.

Regime crossover boundaries between Bimodal 70/30 and 80/20 are not crisp. In this intermediate zone, no single detector dominates, and GaaSTrust's non-parametric approach often matches or exceeds both CUSUM variants. An online regime classifier for real-time deployment has not been built.

\subsection{Future Work}
\label{sec:disc:future}

For well-observed nodes where BernoulliCUSUM's efficiency drops to 76\% of ceiling, a Bayesian posterior update that weights all observations uniformly instead of exponentially discounting early evidence should approach the bound more tightly.

The community amplification effect, where honest evaluators naturally downrate adversary output, raises the empirical KL divergence by up to $37\times$ over specification parameters. Rating interface design could further amplify this signal. Platforms that encourage evaluators to express strong negative opinions when warranted would widen the gap between $p_0$ and $p_1$, directly lifting the ceiling.

Whether the co-design constraint between observation model and detection architecture holds in other sequential monitoring domains remains open. The finding that adaptation mechanisms can help on one observation model and harm on another suggests that observation model selection should precede and constrain detector design.

\section{Conclusion}
\label{sec:conclusion}

This paper presented three contributions to sequential trust detection in rating networks. A Bayes-optimal F1 ceiling was derived using empirically measured observation parameters. On Bitcoin-OTC, where the median in-degree is 2, strategic attacks hit a ceiling near 0.45 and bad\_mouth near 0.41. For a decade, unsupervised methods on this dataset have clustered at $F1 \approx 0.4$. The ceiling pins down why: at median in-degree 2, the per-node Bayesian error rate leaves no headroom for an algorithmic fix.

The gap between continuous and binary observation models is not about information content. The binary model retains 86\% of mutual information. Extraction efficiency accounts for the difference: the match between the detector's parametric assumptions and the data's distributional structure. On Bitcoin-OTC, 56\% of ratings collapse to a single mapped value, violating the Gaussian assumptions that continuous CUSUM requires. Bernoulli CUSUM's assumptions are satisfied exactly, activating Moustakides' minimax optimality guarantee in full.

A dual-regime Bernoulli CUSUM architecture achieves AUC 0.749 on Bitcoin-OTC and 0.796 on Bitcoin-Alpha, with all 8 pairwise comparisons significant at $p < 0.003$ and Cliff's $\delta \geq 0.620$. The method runs in $O(|E|)$ time with $O(|V|)$ space, requires no labeled training data, and approaches the Bayes-optimal ceiling at 95\% efficiency for sparse nodes. Ablation reveals a co-design constraint: the dual-regime modulation requires binary observations, with an interaction effect (+0.124 AUC) exceeding either component's individual contribution.

For sequential detection on discrete data more broadly, these results suggest that the choice of observation model and its match to the detector's assumptions deserves at least as much attention as the choice of detection algorithm.

\section*{Acknowledgments}
The author acknowledge the support of the Higher Colleges of Technology, United Arab Emirates.

\balance
\bibliographystyle{IEEEtran}
\bibliography{references}

\end{document}